\documentclass[preprint]{aastex}

\usepackage{epsfig}

\newcommand{\rremail}{ramaty@gsfc.nasa.gov}
\newcommand{\bzkemail}{benz@wise1.tau.ac.il}
\newcommand{\vtemail}{tatische@csnsm.in2p3.fr}
\newcommand{\nmemail}{natalie@pair.gsfc.nasa.gov}

\slugcomment{ApJL, in press}

\def\lsim{\lower.5ex\hbox{$\; \buildrel < \over \sim \;$}}
\def\gsim{\lower.5ex\hbox{$\; \buildrel > \over \sim \;$}}

\begin{document}

\title{Lithium-6 from Solar Flares}

\author{Reuven Ramaty}
\affil{Laboratory for High Energy Astrophysics\\ NASA/GSFC,
Greenbelt, MD 20771\\ \rremail}

\author{Vincent Tatischeff and J. P. Thibaud}
\affil{Centre de Spectrom\'etrie Nucl\'eaire et de Spectrom\'etrie
de Masse\\ IN2P3-CNRS, 91495 Orsay, France\\ \vtemail}

\author{Benzion Kozlovsky} \affil{School of Physics and Astronomy,
Tel Aviv University, Israel\\ \bzkemail}

\and

\author{Natalie Mandzhavidze} \affil{Laboratory for High Energy Astrophysics, NASA/GSFC\\
and USRA, Greenbelt, MD 20771\\ \nmemail }

\begin{abstract}

By introducing a hitherto ignored $^6$Li producing process, due to
accelerated $^3$He reactions with $^4$He, we show that accelerated
particle interactions in solar flares produce much more $^6$Li
than $^7$Li. By normalizing our calculations to gamma-ray data we
demonstrate that the $^6$Li produced in solar flares, combined
with photospheric $^7$Li, can account for the recently determined
solar wind lithium isotopic ratio, obtained from measurements in
lunar soil, provided that the bulk of the flare produced lithium
is evacuated by the solar wind. Further research in this area
could provide unique information on a variety of problems,
including solar atmospheric transport and mixing, solar convection
and the lithium depletion issue, and solar wind and solar particle
acceleration.

\end{abstract}


\section{Introduction}

The solar wind lithium isotopic ratio, ($^6$Li/$^7$Li)$_{\rm sw}$
=0.032$\pm$0.004, has recently been determined from measurements
in lunar soil (Chaussidon \& Robert 1999). As these authors point
out, this value greatly exceeds the expected photospheric ratio,
based on the fact that $^7$Li in the photosphere is depleted by
over a factor of 100 relative to its protosolar value (i.e. the
photospheric vs. the meteoritic abundance, Grevesse, Noels, \&
Sauval 1996), and that this depletion, due to burning at the
bottom of the convection zone (Brun, Turck-Chieze, \& Zahn 1999),
should lead to a much more severe depletion of $^6$Li, which burns
at a lower temperature than $^7$Li. In addition, there exist
observational upper limits on the photospheric ratio,
($^6$Li/$^7$Li)$_{\rm ph}$$\le$0.01 (M\"uller, Peytremann, \& de
la Reza 1975) and ($^6$Li/$^7$Li)$_{\rm ph}$$\le$0.03 (Ritzenhoff,
Schr\"oter, \& Schmidt 1997). Chaussidon \& Robert (1999) thus
suggest that the measured solar wind $^6$Li must be solar flare
produced. However, they only consider $^6$Li production by
spallation from C, N and O. The demonstration that solar flares
can indeed account for the $^6$Li in the solar wind has very
important implications on many problems in solar physics.

Light element production by accelerated particle interactions was
treated in detail (e.g. Ramaty et al. 1997). In non-solar
settings, and for accelerated particles of predominantly low
energy, the dominant reactions are $^4$He($\alpha$,p)$^7$Li,
$^4$He($\alpha$,n)$^7$Be (with $^7$Be decaying to $^7$Li) and
$^4$He($\alpha$,x)$^6$Li (where x stands for either a proton and a
neutron, or a deuteron). In solar flares, however, the reaction
$^4$He($^3$He,p)$^6$Li is also very important (Mandzhavidze,
Ramaty, \& Kozlovsky 1997a), both because of its very low
threshold energy and because for solar energetic particles
$^3$He/$^4$He can be as large as 1 or even larger (e.g. Reames
1998). Such $^3$He/$^4$He enhancements are one of the main
characteristics of the acceleration mechanism responsible for
impulsive solar energetic particle events, as distinguished from
gradual events, based on the duration of the accompanying soft
X-ray emission. The $^3$He enrichment is thought to be due to
stochastic acceleration through gyroresonant wave particle
interactions which preferentially accelerate the $^3$He (Temerin
\& Roth 1992; Miller \& Vi\~nas 1993). Concerning the particles
which interact at the Sun, evidence for accelerated $^3$He
enrichment was obtained from the detection (Share \& Murphy 1998)
of a gamma-ray line at 0.937 MeV produced by the reaction
$^{16}$O($^3$He,p)$^{18}$F$^*$ (Mandzhavidze, Ramaty, \& Kozlovsky
1997b; 1999). Using gamma-ray data from 20 flares, Mandzhavidze et
al. (1999) showed that for essentially all of these flares
$^3$He/$^4$He can be as large as 0.1, while for some of them
values as high as 1 are possible. In addition, they showed that
for the particles that interact and produce gamma rays, $^3$He
enrichments are present for both impulsive and gradual flares.
Thus, we can expect $^3$He/$^4$He$\gsim$0.1 for most flares that
produce gamma rays and isotopes at the Sun.

In the present Letter we carry out new calculations of Li
production and  re-calculate (see Ramaty \& Simnett 1991) the
average accelerated ion irradiation of the Sun, to show that flare
accelerated particle interactions produce enough $^6$Li which,
combined with photospheric $^7$Li, can account for the solar wind
$^6$Li/$^7$Li measured in lunar soil.

\section{Li Production}

We employ the nuclear code described in detail in Ramaty et al.
(1997) which includes, in addition to the $\alpha$$\alpha$
reactions mentioned above, also Li production from C, N and O. The
cross section for the additional reaction, $^4$He($^3$He,p)$^6$Li,
is shown in Figure~1, together with the cross sections for the
$\alpha$$\alpha$ reactions producing $^6$Li and $^7$Li. For the
$^3$He induced reaction we obtained the cross section for $^6$Li
production in the ground state, from threshold (2.34 MeV/nucleon)
to 8.2 MeV/nucleon, by detailed balance using the cross section
for the inverse exothermic reaction $^6$Li(p,$^3$He)$^4$He (Angulo
et al. 1999). We added the contribution of the reaction for
producing $^6$Li in the 3.56 MeV excited state which decays to the
ground state by photon emission, using data from Harrison (1967).
The total cross section at 9.3 MeV/nucleon is from Koepke and
Brown (1977), and at 18 and 20.4 MeV/nucleon from Halbert, van der
Woude, \& O'Fallon (1973). At higher energies we extrapolated the
cross section as expected for reactions with 2 particles in the
exit channel.

Gamma-ray production in solar flares results predominantly from
thick target interactions, meaning that particles accelerated in
the upper portions of coronal loops produce nuclear reactions as
they slow down in the denser chromospheric region of the loops
(e.g. Ramaty \& Murphy 1987). We adopt the same model for Li
production. The upper panel in Figure~2 shows the resultant thick
target $^6$Li yields, normalized to unit incident total number of
protons of energy greater than 30 MeV, $N_{\rm p}(>$30)=1. The
energy spectra of the accelerated particles are power laws in
kinetic energy per nucleon, with spectral index $s$ (Ramaty,
Mandzhavidze, \& Kozlovsky 1996). The evidence for enhanced
$^3$He/$^4$He was mentioned above. There is also evidence that
$\alpha$/p could exceed the canonical 0.1, with possible value
around 0.5 (Share \& Murphy 1997; Mandzhavidze et al. 1999). Thus
in Figure~2 we show results for $\alpha$/p = 0.1 and 0.5, and
$^3$He/$^4$He=0, 0.1 and 1. We see in the upper panel that the
$^3$He enrichment very significantly increases the lithium
production, especially for steep spectra. That the lithium
production is mainly due to $\alpha$ particles and $^3$He nuclei
can be seen by comparing the six upper curves with the lowest one,
for which we set the $\alpha$ particle and $^3$He abundances to
zero, so that all the $^6$Li in this case is produced in C, N and
O interactions. Considering the flare produced isotopic ratios in
the lower panel, we see that while in the absence of $^3$He,
$^6$Li/$^7$Li is at most unity, much larger ratios are possible
with enhanced $^3$He/$^4$He.

\section{Average Solar Proton Irradiation}

To calculate the average flare produced lithium, we estimate the
average proton irradiation of the Sun, $\dot N_{\rm p}$($>$30MeV)
measured in protons per second, where the average is taken over a
solar cycle. We follow the method described by Ramaty \& Simnett
(1991). We start with the flare size distribution measured in 0.3
to 1 MeV bremsstrahlung because observations in this energy range
give the most complete sample of solar flare gamma-ray emission
(see Vestrand et al. 1999). To minimize the effects of anisotropic
electrons (e.g. Miller \& Ramaty 1989) we employ the distribution
derived for flares near the solar limb (Dermer 1987). For flares
at heliocentric longitudes 60$^\circ$ to 90$^\circ$, observed from
March 1980 to February 1986 (approximately half a solar cycle) the
size distribution, measured in number of flares per unit $F_{\rm
B}$, can be approximated by $dn/dF_{\rm B}\simeq 8.5 F_{\rm
B}^{-1.1}$, where 10$\lsim F_{\rm B}\lsim$6500 photons cm$^{-2}$
is the observed 0.3 to 1 MeV bremsstrahlung fluence at Earth per
flare . The total number of $>$0.3 MeV emitting flares per solar
cycle is obtained by integrating the above expression multiplied
by a factor of 12, where a factor of 6 takes into account the
whole solar surface and a factor of 2 the other half of solar
cycle. We thus obtain 375 flares, which compares well with the 175
flares listed by Vestrand et al. (1999) from which $>$0.3 MeV
bremsstrahlung was observed with the Solar Maximum Mission (SMM)
over almost a whole solar cycle. This latter number should be
corrected for anisotropy effects, and must be multiplied by a
factor of 2 since SMM only observes half the solar surface. The
required average irradiation is then given by
\begin{eqnarray}
\dot N_{\rm p}(>30) = {12\over T} \int_{10}^{6500} dF_{\rm B}
{dn\over dF_{\rm B}} N_{\rm p}(F_{\rm B})~,
\end{eqnarray}
where T is the number of seconds in 11 years and $N_{\rm p}(F_{\rm
B })$ is the number of protons above 30 MeV expressed as a
function of $F_{\rm B}$. To derive this relationship, we first
employ the result of Murphy et al. (1990) that for flares near the
limb $F_{\rm B}/F_{\rm N} \simeq 4.5$, where $F_{\rm N}$ is the
total nuclear deexcitation line emission fluence observed at
Earth. Next we use the nuclear deexcitation code (e.g. Ramaty et
al. 1996) to derive $N_{\rm P}(>$30)/$F_{\rm N}$. This ratio
depends on the spectrum and composition of the accelerated
particles, in particular $\alpha$/p. Ramaty et al. (1996) have
derived the distribution of power law spectral indexes from
gamma-ray data, showing that for a sample of 19 flares the mean
$s\simeq 4$. For this value of $s$ we find that $N_{\rm
p}(>$30)/$F_{\rm N}$ = 1.7$\times$10$^{29}$ and
6.6$\times$10$^{29}$ protons/(nuclear deexcitation photons
cm$^{-2}$), for $\alpha$/p=0.5 and 0.1, respectively. By using
$F_{\rm B}/F_{\rm N}$=4.5 and these $N_{\rm p}(>$30)/$F_{\rm N}$
to derive $N_{\rm p}(F_{\rm B})$, equation~1 yields $\dot N_{\rm
p}$($>$30MeV)=3.5$\times$10$^{25}$ and 1.4$\times$10$^{26}$
protons s$^{-1}$, for $\alpha$/p=0.5 and 0.1, respectively.

\section{The Solar Wind $^6$Li/$^7$Li}

Even though a detailed treatment of the time dependent evolution
of Li in the solar atmosphere is beyond the scope of this paper,
we now show that $^6$Li production in solar flares could indeed
account for the solar wind $^6$Li/$^7$Li. To demonstrate this we
assume the following: (i) all the flare produced $^6$Li is
evacuated by the solar wind, (ii) the photospheric $^6$Li that is
the remnant of its protosolar abundance is negligible, and (iii)
the solar wind ($^7$Li/H)$_{\rm sw}$ is equal to the photospheric
value ($^7$Li/H)$_{\rm ph}$=1.4$\times$10$^{-11}$ (Grevesse et al.
1996). The solar wind ($^6$Li/$^7$Li)$_{\rm sw}$ is then given by
\begin{eqnarray}
\Big({^6{\rm Li} \over ^7{\rm Li}}\Big)_{\rm sw}= {\dot N_{\rm
p}(>30) Q(^6{\rm Li})/ \dot F_{\rm sw} \over (^7{\rm Li}/{\rm
H})_{\rm ph}}~,
\end{eqnarray}
where $Q(^6{\rm Li})$ is plotted in Figure~2 and $\dot F_{\rm
sw}$$\simeq$6$\times$10$^{35}$ s$^{-1}$ is the average solar wind
proton flux (Dupree 1996). Taking $s=4$, 0.1$<$$\alpha$/p$<$0.5
and 0.1$<$$^3$He/$^4$He$<$1, we obtain
0.007$<$($^6$Li/$^7$Li)$_{\rm sw}$$<$0.06. This range is
consistent with the observed value of 0.032$\pm$0.004. Several
effects could lead to lower or higher calculated
($^6$Li/$^7$Li)$_{\rm sw}$. Clearly there are uncertainties in our
estimate of $\dot N_{\rm p}(>$30 MeV), in particular there could
be a large number of smaller gamma-ray flares, which have not yet
been observed, and if they had steep ion energy spectra and high
$^3$He/$^4$He they would contribute significantly to $^6$Li
production. On the other hand, some of the flare-produced $^6$Li
could be mixed downward to the photosphere and lost from the solar
wind. The calculated ($^6$Li/$^7$Li)$_{\rm sw}$ would also be
lower if ($^7$Li/H)$_{\rm sw}$ were higher than ($^7$Li/H)$_{\rm
ph}$, a possibility since Li has low first ionization potential, a
factor that biases coronal abundances relative to those of the
photosphere (e.g. Reames 1998). Nevertheless, the better than
order of magnitude agreement between the calculated and measured
($^6$Li/$^7$Li)$_{\rm sw}$ provides good support to the
possibility that the measured $^6$Li in lunar soil is indeed solar
flare produced.

It is of some interest to compare the average $^6$Li production,
6$\times$10$^{22}$$<$[$\dot N_{\rm p}$$(>$30)$Q(^6$Li)]
$<$5$\times$10$^{23}$ atoms s$^{-1}$, with the contribution of the
19 large SMM flares from which gamma-ray line emission was
observed. Using the method detailed in Mandzhavidze et al. (1999),
for each flare we derive $s$ and $N_{\rm p}$($>$30). Then using
$Q$($^6$Li) from Figure~2, taking into account that for 5 of the
19 flares $\alpha$/p$\simeq$0.5 (Mandzhavidze et al. 1999), we
obtain the flare-by-flare $^6$Li productions which yield averages
over the 9 year SMM observing period of 1$\times$10$^{22}$ and
7$\times$10$^{22}$ $^6$Li atoms s$^{-1}$, if for all flares
$^3$He/$^4$He=0.1 and 1, respectively. Thus, about 15\% of the
$^6$Li production that we derived using the flare size
distribution could result from 19 of the largest flares.
Concerning the contributions of individual flares, as much as a
few times 10$^{30}$ Li atoms could be produced by a large flare
and most of these would be $^6$Li (Mandzhavidze et al. 1997a)

\section{Discussion and Conclusions}

We demonstrated that it is possible to produce enough $^6$Li by
flare accelerated particles to account for the measured
$^6$Li/$^7$Li in lunar soil that is thought to originate from
solar wind implantation. The presence of enriched accelerated
particle $^3$He is essential for the production of sufficient
$^6$Li. We note that the radioactive $^{26}$Al in the early solar
system is thought to be produced in $^3$He induced reactions (Lee
et al. 1998). This raises the possibility that some of the
meteoritic $^6$Li could also be of local early solar system
origin.

Kotov et al. (1996) claimed that flare accelerated particle
interactions could account for all the photospheric lithium. If
this were true, since the solar wind acceleration is not expected
to significantly alter the lithium isotopic ratio, the solar wind
$^6$Li/$^7$Li should exceed 0.2 (Figure~2), contrary to the
observed value of 0.03. This confirms the previous result of
Mandzhavidze et al. (1997a) that production in flares does not
make a significant contribution to the average photospheric
lithium. But the fact that as much as 10$^{30}$ Li atoms are
produced in large solar flares, suggests that flare produced
lithium may be detected in a small area of the solar surface near
the foot points of the flaring loops shortly after the time of the
flare (see Livshits 1997). In this connection, it is interesting
to point out that Ritzenhoff et al. (1997) don't rule out the
presence of $^6$Li near a sunspot at a value close to their
reported upper limit $^6$Li/$^7$Li$\le$0.03, which in fact
coincides with the measured solar wind value.

Further research in this area requires direct measurement of
lithium and its isotopic ratio in the solar wind, spectroscopic
measurements of $^6$Li in the photosphere, and the detection of
gamma rays from small flares that would lead to a more precise
determination of the proton irradiation of the Sun. All of these
should lead to new insights into the processes of transport and
mixing in the solar atmosphere and of the acceleration of the
solar wind.

\eject

\begin{figure}[t]
  \begin{center}
    \leavevmode
\epsfxsize=15.cm \epsfbox{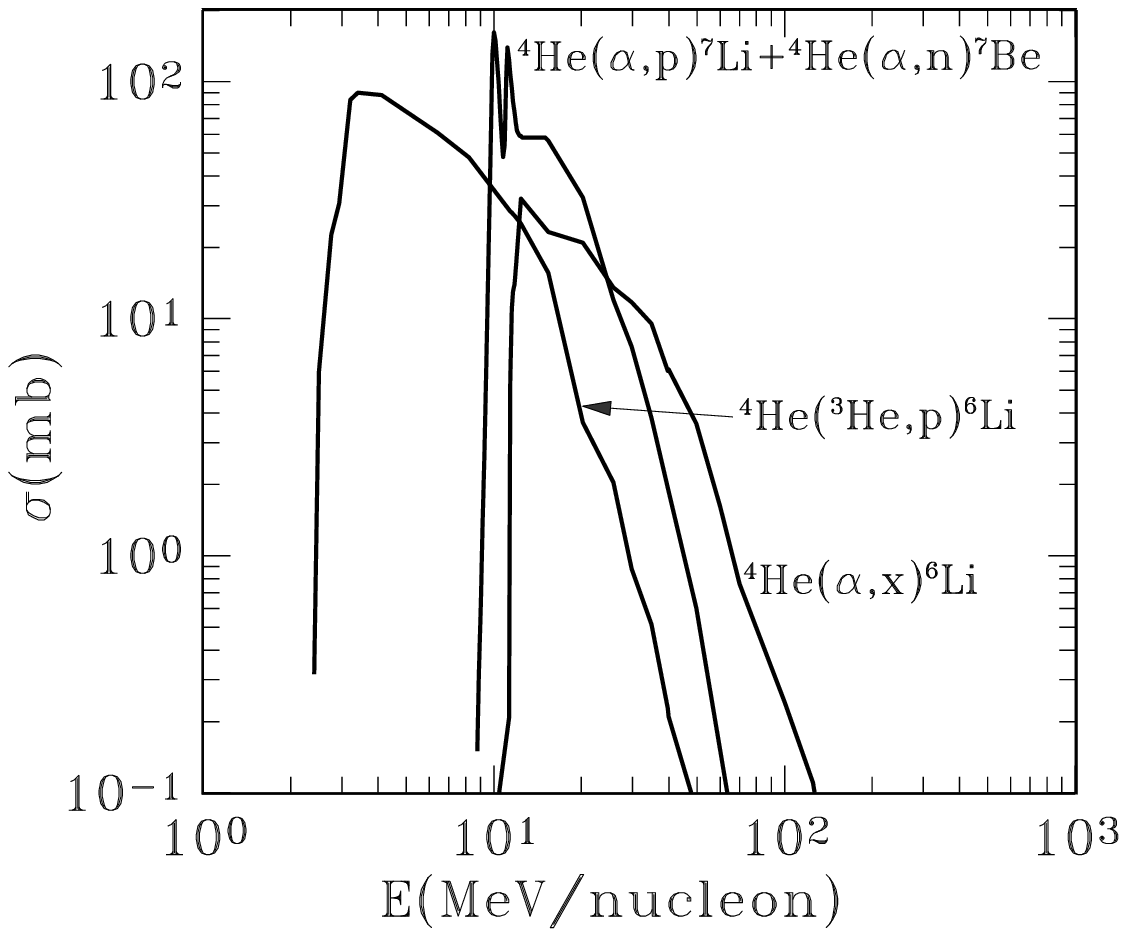}
\end{center}
\caption{$^6$Li production cross sections in accelerated $^3$He
and $\alpha$ particle interactions with He.}
\end{figure}

\begin{figure}[t]
  \begin{center}
    \leavevmode
\epsfxsize=11.cm \epsfbox{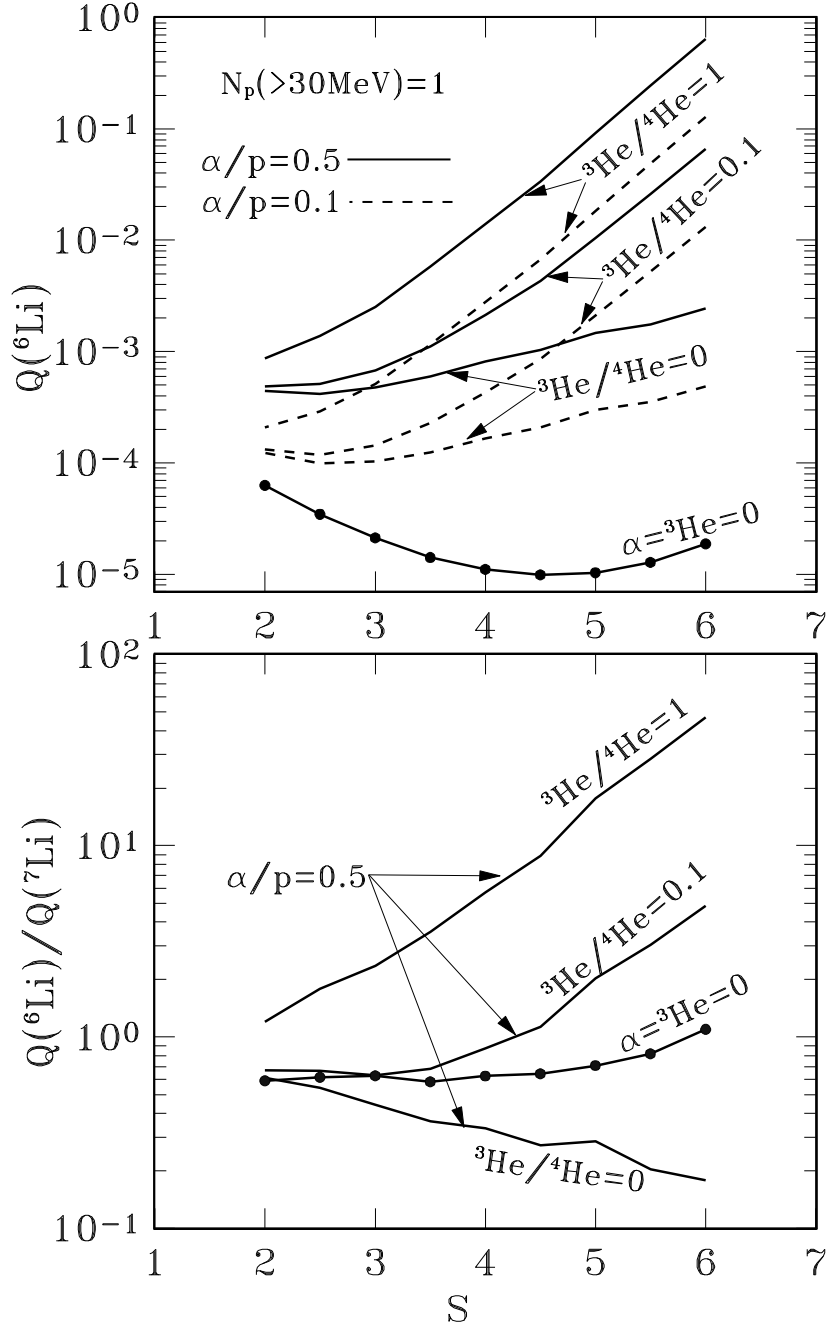}
\end{center}
\caption{Upper panel: Thick target $^6$Li productions by
accelerated particles with power law in kinetic energy per nucleon
spectra with spectral index $s$ and normalized to 1 proton of
energy greater than 30 MeV. For the curve with $\alpha$=$^3$He=0,
the production is due solely to CNO interactions. Lower panel:
Isotopic ratios; for $\alpha$/p=0.1 (not shown) the values are
practically identical to those shown for $\alpha$/p=0.5.}
\end{figure}

\end{document}